\documentclass[conference]{IEEEtran}
\IEEEoverridecommandlockouts

\usepackage{cite}
\usepackage{amsmath,amssymb,amsfonts}
\usepackage{algorithmic}
\usepackage{graphicx}
\usepackage{textcomp}
\usepackage{booktabs,multirow,colortbl,xcolor}
\usepackage[table]{xcolor}
\usepackage{cleveref}
\usepackage{url}
\def\BibTeX{{\rm B\kern-.05em{\sc i\kern-.025em b}\kern-.08em
    T\kern-.1667em\lower.7ex\hbox{E}\kern-.125emX}}

\begin{document}

\title{Language Orthogonalization of\\Self-Supervised Speech Representations for\\Cross-lingual Parkinson's Detection
}
\author{
  \IEEEauthorblockN{
    Minu Kim\textsuperscript{1},
    Eunjung Yeo\textsuperscript{2},
    Kwanghee Choi\textsuperscript{2},
    June-Woo Kim\textsuperscript{3}
  }
  \IEEEauthorblockA{
    \textsuperscript{1}University of Southern California, USA \,\, \textsuperscript{2}The University of Texas at Austin, USA  \,\, \textsuperscript{3}Wonkwang University, South Korea\\
    minukim@usc.edu \,\, \{eunjung.yeo,kwanghee\}@utexas.edu \,\, kaen2891@wku.ac.kr 
  }
}

\maketitle

\begin{abstract}
Self-supervised speech models (S3Ms) provide powerful representations for Parkinson's disease (PD) detection, making cross-lingual transfer attractive for languages lacking labeled patient speech. However, these representations also encode language identity, which can confound this transfer: without target-language PD speech, classifiers may separate languages rather than pathology, yielding high specificity but low sensitivity on target patients. We propose \emph{language orthogonalization}, a closed-form ridge residualization of S3M features against external VoxLingua107 language embeddings, fitted using only healthy-control (HC) speech. By removing language-predictable components while retaining pathology-related variation, it produces a less language-dependent geometry in which HC representations concentrate while PD representations disperse. Across five S3M backbones, three speech tasks, and three target languages, our method consistently improves cross-lingual PD-detection performance while correcting the high-specificity/low-sensitivity failure.\footnote{Source code: \url{https://github.com/MINUKIMS/language-orthogonalization}.}
\end{abstract}
\begin{IEEEkeywords}
self-supervised speech representation, dysarthric speech, parkinson's disease, cross-lingual processing.
\end{IEEEkeywords}

\begin{figure*}[t]
    \centering
    \includegraphics[width=0.9\textwidth]{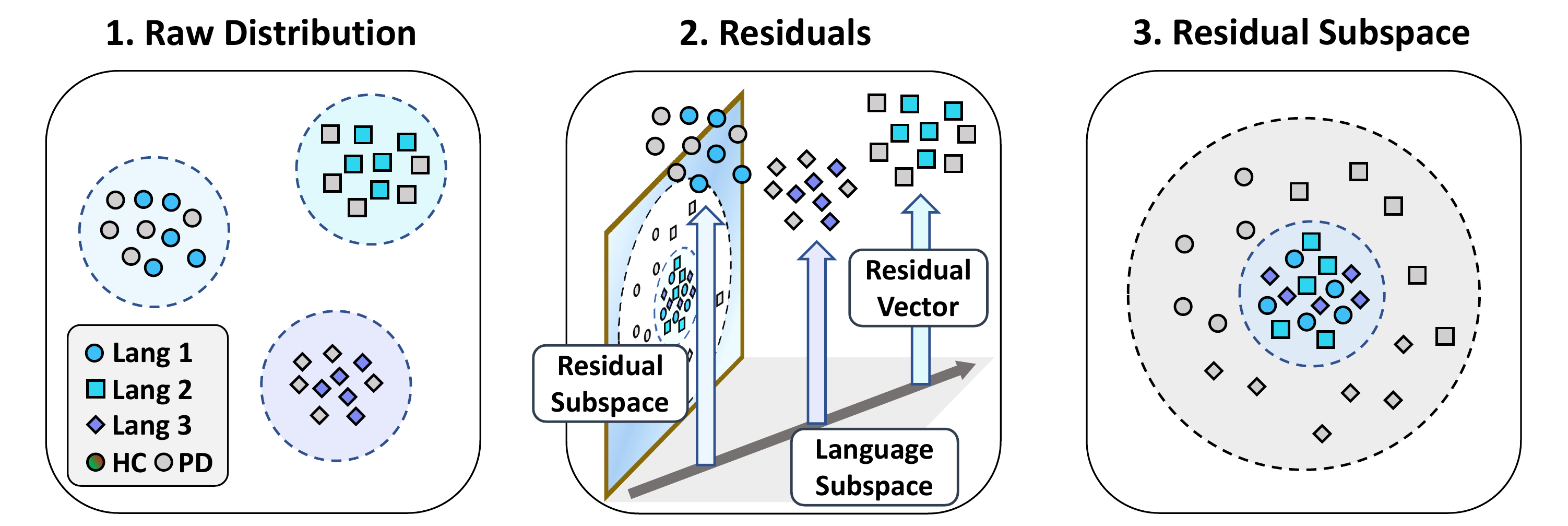}
    \caption{\textbf{Schematic of language orthogonalization.} Raw representations are orthogonally projected onto a language subspace. Extracting the residual vectors yields a language-orthogonal subspace that effectively isolates the underlying pathology signal. Ideally, Healthy Controls (HC) stably cluster together in this space, while Parkinson's Disease (PD) representations scatter outward, reflecting the sporadic and heterogeneous nature of the disease.}
    \label{fig:orthogonalization}
\end{figure*}

\section{Introduction}

Parkinson’s disease (PD) is a progressive neurodegenerative disorder that commonly impairs motor speech, most characteristically through hypokinetic dysarthria~\cite{critchley1981speech,atalar2023hypokinetic}. Its effects span phonation, articulation, and prosody, manifesting as reduced loudness, monotonic pitch, imprecise articulation, altered timing, and diminished intelligibility~\cite{darley1969differential,duffy2012motor,kovac2025digital,cao2025speech}. Because these impairments can be captured non-invasively, repeatedly, and at low cost, speech-based digital biomarkers are increasingly being investigated as scalable indicators of motor-speech dysfunction~\cite{dibazar2002feature,dibazar2006pathological,gupta2016pathological,rusz2024prodromal}.

Despite this promise, most speech-based PD detection systems have been developed and evaluated within a single language or corpus~\cite{tsanas2012novel,kim2015automatic,bocklet2011detection,thies2025automatic}. This limitation is particularly important for speech-based assessment because observed speech patterns reflect not only PD-related impairment but also language-specific phonology, syllable structure, rhythm, and recording conditions~\cite{orozco2016automatic,kim2017cross,lim2025cross,kim2025cross,yeo2022multilingual,yeo2025applications}. Cross-lingual PD detection must therefore distinguish pathology-related characteristics that transfer across languages from linguistic and corpus-specific variation that does not. Self-supervised speech models (S3Ms) offer a promising foundation for this problem: pre-trained on large-scale speech data, they encode rich acoustic and phonetic information~\cite{hsu2021hubert,chen2022wavlm,babu2022xls} and have outperformed conventional handcrafted features in dysarthria and PD detection~\cite{yeo2023automatic,javanmardi2024pre,sapkota2025all}.

The linguistic richness of S3Ms, however, poses a critical challenge for cross-lingual PD detection: these models encode strong language-dependent structure alongside pathology-relevant cues~\cite{hernandez2026adapting}. Consider training with healthy control (HC) and PD speech from resource-rich source languages but only HC speech from the target language—a realistic setting when target-language patients are unavailable~\cite{li2025towards,yeo2025applications,hernandez2026adapting}. Because all target-language training speakers are HC, the classifier may associate the target language with HC rather than learn transferable pathology cues. As a result, unseen target-language patients tend to be misclassified as HC. On the OneVoice-MSD-2026 benchmark, this bias produces high specificity but low sensitivity: healthy speakers are rarely misclassified, whereas many patients are missed~\cite{hernandez2026adapting}. This failure is particularly problematic for screening, where high sensitivity is needed to minimize false negatives~\cite{maxim2014screening,bone2016use}.\footnote{Sensitivity, or the true-positive rate, is the proportion of patients correctly identified; specificity, or the true-negative rate, is the proportion of healthy controls correctly identified.}

Prior work has shown that S3Ms encode phonetic context in an approximately linear structure across diverse languages~\cite{choi2026self,choi2026b}. In cross-lingual PD detection, language shift (LS)~\cite{hernandez2026adapting} adopts a related geometric view, representing language differences as linear vectors that align the HC centroids of the source and target languages. Because LS translates every speaker within a language by the same vector, it changes the distribution mean but not its within-language covariance, allowing language-dependent structure beyond the centroid to persist.

We extend this view by estimating language-dependent variation for each speaker. Specifically, we propose \emph{language orthogonalization} (\Cref{fig:orthogonalization}), which fits an HC-only ridge regression from external language-identification (LID) embeddings~\cite{valk2021voxlingua107} to S3M representations and subtracts the predicted component. Because this component varies across speakers rather than being fixed for each language, it can reduce language-dependent structure beyond the distribution mean. Fitting only on HC speech prevents the regression from directly modeling PD-associated variation, thereby reducing the risk of suppressing pathology-related deviations.

Our contributions are threefold. $(i)$ We introduce \emph{language orthogonalization}, an HC-only closed-form transform that reduces language-predictable variation without target-language PD speech. $(ii)$ We demonstrate consistent improvements across five S3M backbones, three speech tasks, and three target languages, including the transfer of sensitivity-selected thresholds to unseen target-language patients. $(iii)$ Through language probing, we show that the resulting residuals substantially suppress language decodability.

\section{Method}
\label{sec:method}

Our pipeline has three stages before downstream classification (\Cref{fig:orthogonalization}). First, a frozen self-supervised speech model (S3M) extracts a layer-wise pooled utterance embedding (\Cref{sec:pooling}). Second, an external language-identification (LID) model provides a language reference embedding (\Cref{sec:lid}). Third, an alignment step subtracts the S3M component predictable from the LID embedding and retains the residual (\Cref{sec:methods-align}). This residual is expected to reduce language separation while retaining pathology-related variation.

\subsection{S3M backbones and pooling}
\label{sec:pooling}
We use frozen S3Ms as utterance-level feature extractors. To probe whether the alignment step is robust to the choice of representation, we span five 24-layer Transformer-Large encoders (1024-d hidden states): wav2vec2-Large-LV60~\cite{baevski2020wav2vec} (English, 60k\,h), XLS-R-300M~\cite{babu2022xls} (128 languages, 436k\,h), MMS-300M~\cite{pratap2024scaling} (1162 languages, 491k\,h), WavLM-Large~\cite{chen2022wavlm} (English, 94k\,h), and HuBERT-Large~\cite{hsu2021hubert} (English, 60k\,h). From each \emph{frozen} encoder we compute mean+std pooling~\cite{klempir2026statistical} over each of the 25 representations (convolutional output plus 24 transformer layers), $\ell_2$-normalize per layer, and average across layers into a 2048-d utterance embedding. Per-speaker representations are obtained by averaging the utterance embeddings within each task. Because information in S3Ms is distributed across layers and varies by backbone~\cite{pasad2021layer,pasad2023comparative}, and for PD detection the most informative depth is not settled~\cite{klempir2026statistical,la2024exploiting}, we average across all depths rather than tune a layer choice. This pipeline is fixed across all backbones, so any downstream difference isolates the alignment step.

\subsection{External LID embeddings}
\label{sec:lid}

To provide an external reference for language-dependent variation, we use the VoxLingua107 ECAPA-TDNN~\cite{valk2021voxlingua107}, a 14M-parameter encoder pre-trained to identify 107 languages. Its embeddings emphasize language-discriminative information without relying on PD labels. We extract a 256-d embedding for each utterance and average the embeddings within each speaker $\times$ task to obtain a speaker-level LID vector $g_i\in\mathbb{R}^{256}$, where the tasks are defined in \Cref{ssec:dataset}.

\subsection{Alignment methods}
\label{sec:methods-align}

Let $x_i\in\mathbb{R}^{D}$ denote the speaker-level S3M representation of speaker $i$, and let $X=[x_1^\top,\ldots,x_N^\top]^\top\in\mathbb{R}^{N\times D}$ stack all $N$ representations, with $D=2048$. Let $\ell_i$ denote speaker $i$'s language, $T$ the target language, and $\mathcal{H}_L$ the training-fold HC indices for language $L$, with $\mathcal{H}=\bigcup_L\mathcal{H}_L$. Both alignment methods are fitted on $\mathcal{H}$ and applied to all training and evaluation speakers.
 
\emph{Language shift} (LS)~\cite{hernandez2026adapting} applies a constant offset to each language, aligning its training-fold HC centroid $\mu_{\ell_i}$ with the target-language HC centroid $\mu_T$:
\begin{equation}
x_i' = x_i-\mu_{\ell_i}+\mu_T,
\qquad
\mu_L=\frac{1}{|\mathcal{H}_L|}
\sum_{j\in\mathcal{H}_L}x_j .
\label{eq:ls}
\end{equation}

\emph{Language orthogonalization} (LO; proposed) estimates the S3M component linearly predictable from an LID embedding. Let $g_i\in\mathbb{R}^{256}$ denote speaker $i$'s LID vector, and let $X_{\mathcal H}$ and $G_{\mathcal H}$ stack the corresponding S3M and LID representations for $\mathcal H$. We fit a ridge map $W^*\in\mathbb{R}^{D\times256}$ on this subset:

\begin{align}
W^*
&= \operatorname*{argmin}_{W}
\sum_{i\in\mathcal H}
\lVert x_i-Wg_i\rVert_2^2
+\alpha\lVert W\rVert_F^2 \\
&= \operatorname*{argmin}_{W}
\lVert X_{\mathcal H}-G_{\mathcal H}W^\top\rVert_F^2
+\alpha\lVert W\rVert_F^2 ,
\label{eq:ridge-obj}
\end{align}

where $\lVert \cdot \rVert_F$ is the Frobenius norm and $\alpha$ is the ridge coefficient, with smaller values removing more LID-predictable variation.
We report results as a function of $\alpha$ (\Cref{sec:downstream}).

The objective in \Cref{eq:ridge-obj} has the following closed-form ridge solution. For each speaker, $W^*g_i$ denotes the S3M component predicted from the LID embedding, and $x_i'$ is the residual used for downstream classification:
\begin{equation}
\begin{aligned}
W^*
&= X_{\mathcal H}^{\top}G_{\mathcal H}
\left(G_{\mathcal H}^{\top}G_{\mathcal H}+\alpha I\right)^{-1},\\
x_i'
&= x_i-W^*g_i .
\end{aligned}
\label{eq:orth}
\end{equation}
Here, $\alpha>0$ is the ridge regularization coefficient, which ensures that the regularized Gram matrix is invertible. Unlike LS, which subtracts one fixed vector per language, LO subtracts a language-predictable component that varies across speakers. It can therefore reduce language-dependent variation beyond the distribution mean rather than merely translate the centroid.

\section{Experimental Setups}\label{sec:setups}
\subsection{Dataset}\label{ssec:dataset}
We use the OneVoice-MSD-2026 corpus~\cite{hernandez2026adapting}, a collaborative benchmark that harmonizes three PD-speech databases under a common protocol: GermanPD ($n{=}176$)~\cite{bocklet2011detection}, CzechPD ($n{=}100$)~\cite{rios2024automatic}, and the Spanish e-PC-GITA ($n{=}140$)~\cite{perez2021emotional}, with $n$ speakers. Each speaker performs three speech tasks~\cite{duffy2012motor}: sustained vowel phonation (a prolonged vowel), oral diadochokinesis (DDK, rapid /pa-ta-ka/ repetition), and read speech (reading full sentences aloud), all labeled binary HC versus PD. We split each language into five speaker-disjoint folds, stratified jointly by group (HC/PD) and language so that HC/PD balance is preserved within every language.

\subsection{Cross-lingual protocol}
We follow the same cross-lingual protocol as in~\cite{hernandez2026adapting}: target pathology is never seen in training, so a classifier cannot rely on memorizing target-language acoustics and must transfer pathology cues from the source languages.
For each target language $T$ and outer fold $k$, training uses all source-language speakers together with the target-language HC folds other than $k$; evaluation uses the held-out target HC fold and \emph{all} target-language PD speakers:
\begin{equation}
\begin{aligned}
\mathcal{D}_{\text{train}} ={}& \{s : \mathrm{lang}(s)\neq T\}\\
&\cup\ \{s : \mathrm{lang}(s){=}T,\ \mathrm{HC},\ \mathrm{fold}(s)\neq k\},\\[2pt]
\mathcal{D}_{\text{eval}} ={}& \{s : \mathrm{lang}(s){=}T,\ \mathrm{HC},\ \mathrm{fold}(s){=}k\}\\
&\cup\ \{s : \mathrm{lang}(s){=}T,\ \mathrm{PD}\}.
\end{aligned}
\nonumber
\end{equation}
On each fold, we fit a PD detection model through class-balanced logistic regression on the pooled representations, where features are standardized on training statistics only, using the same settings of \cite{hernandez2026adapting}.

\subsection{Metrics}
The primary metric for the PD detection model is F1 at the sensitivity-$0.90$ operating point of~\cite{hernandez2026adapting}, reported as the mean over 5 backbones $\times$ 3 tasks $\times$ 3 target languages $\times$ 5 folds (\Cref{sec:downstream}.
Further, to test whether the threshold from the training data still works on the test data, we additionally report the cross-lingual sensitivity \emph{achieved} on the target language against the sensitivity \emph{requested} at training (\Cref{sec:downstream}).

\begin{table}[t]
\centering
\small
\setlength{\tabcolsep}{7pt}
\renewcommand{\arraystretch}{1.2}
\caption{\textbf{Cross-lingual F1 at the 0.90-sensitivity operating point.}
Results are averaged over five S3M backbones and three target languages.
LO results are reported at $\alpha=0$ for all tasks and backbones.}
\label{tab:f1_09}
\begin{tabular}{l cc cc}
\toprule
\multirow{2}{*}{Task}
& \multicolumn{2}{c}{Baselines}
& \multicolumn{2}{c}{LO} \\
\cmidrule(lr){2-3}\cmidrule(lr){4-5}
& \raisebox{0.5\baselineskip}{Raw}
& \raisebox{0.5\baselineskip}{LS}
& \shortstack{F1\\($\alpha=0$)}
& \shortstack{Gain\\over Raw} \\
\midrule
Vowel & 0.53 & 0.65 & \textbf{0.89} & $\mathbf{+0.36}$ \\
DDK   & 0.59 & 0.68 & \textbf{0.90} & $\mathbf{+0.31}$ \\
Read  & 0.43 & 0.69 & \textbf{0.81} & $\mathbf{+0.38}$ \\
\midrule
\textit{Avg.}
& 0.52 & 0.67 & \textbf{0.87} & $\mathbf{+0.35}$ \\
\bottomrule
\end{tabular}
\end{table}
 
\begin{figure}[t]
  \centering
  \includegraphics[width=\columnwidth]{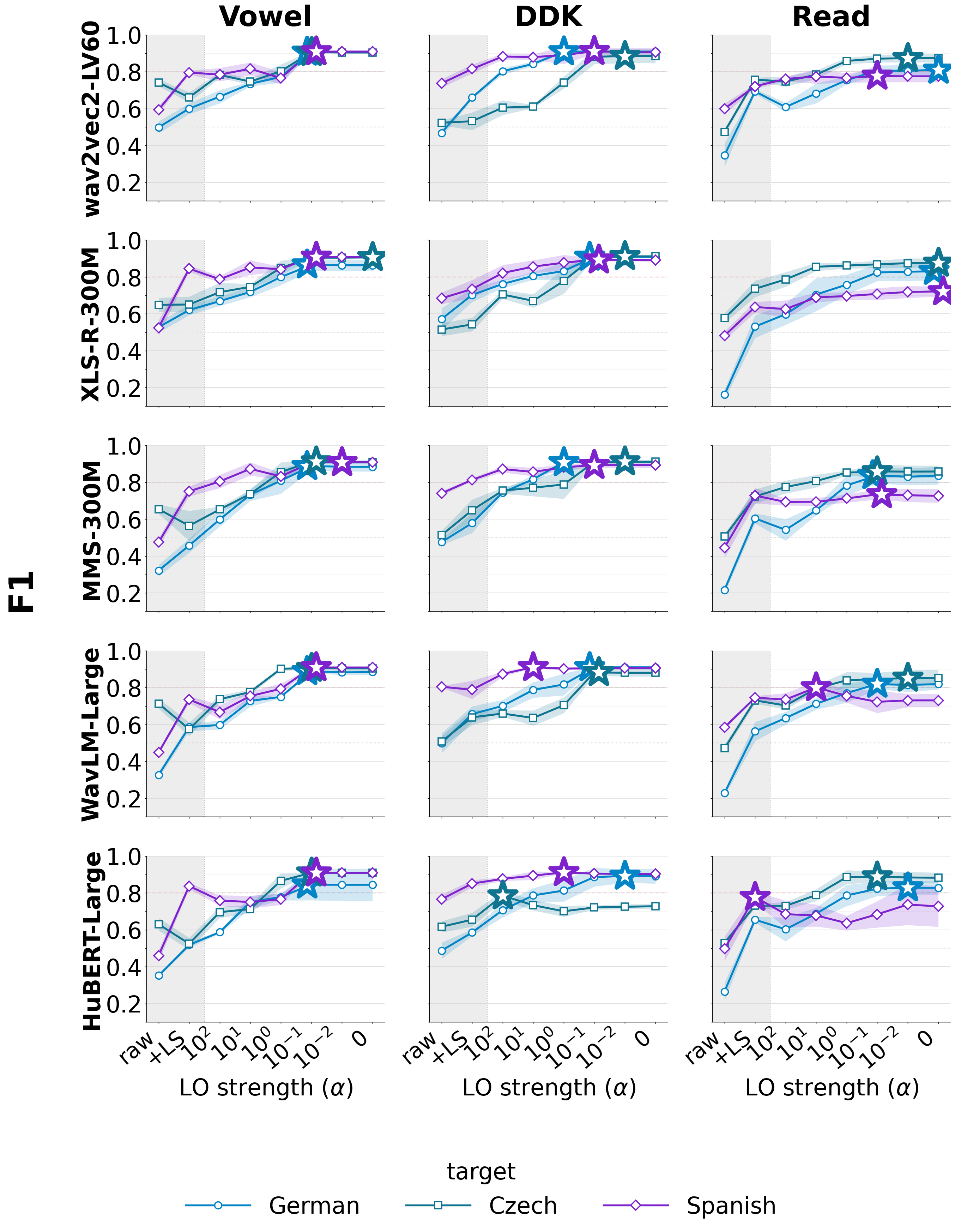}
  \caption{\textbf{Cross-lingual PD-detection F1 versus orthogonalization strength $\alpha$.}
  Results are shown for five S3M backbones (rows) and three speech tasks (columns), averaged over three target languages. The shaded region on the left shows the raw and $+$LS baselines, and stars mark the maximum F1 along each curve. LO outperforms both baselines over a broad range of $\alpha$, with F1 generally increasing as $\alpha$ decreases.}
  \label{fig:f1_alpha}
\end{figure}

\section{Results}\label{sec:downstream}
We now test whether our method improves cross-lingual PD detection over the baselines in \Cref{ssec:sota}.
We additionally focus on the high-specificity, low-sensitivity problem \cite{hernandez2026adapting}.
We examine whether the detection thresholds are transferable from the train to test settings in \Cref{ssec:thr}, and explore the application to PD screening in \Cref{ssec:screen}.

\begin{figure}[t]
  \centering
  \includegraphics[width=\columnwidth]{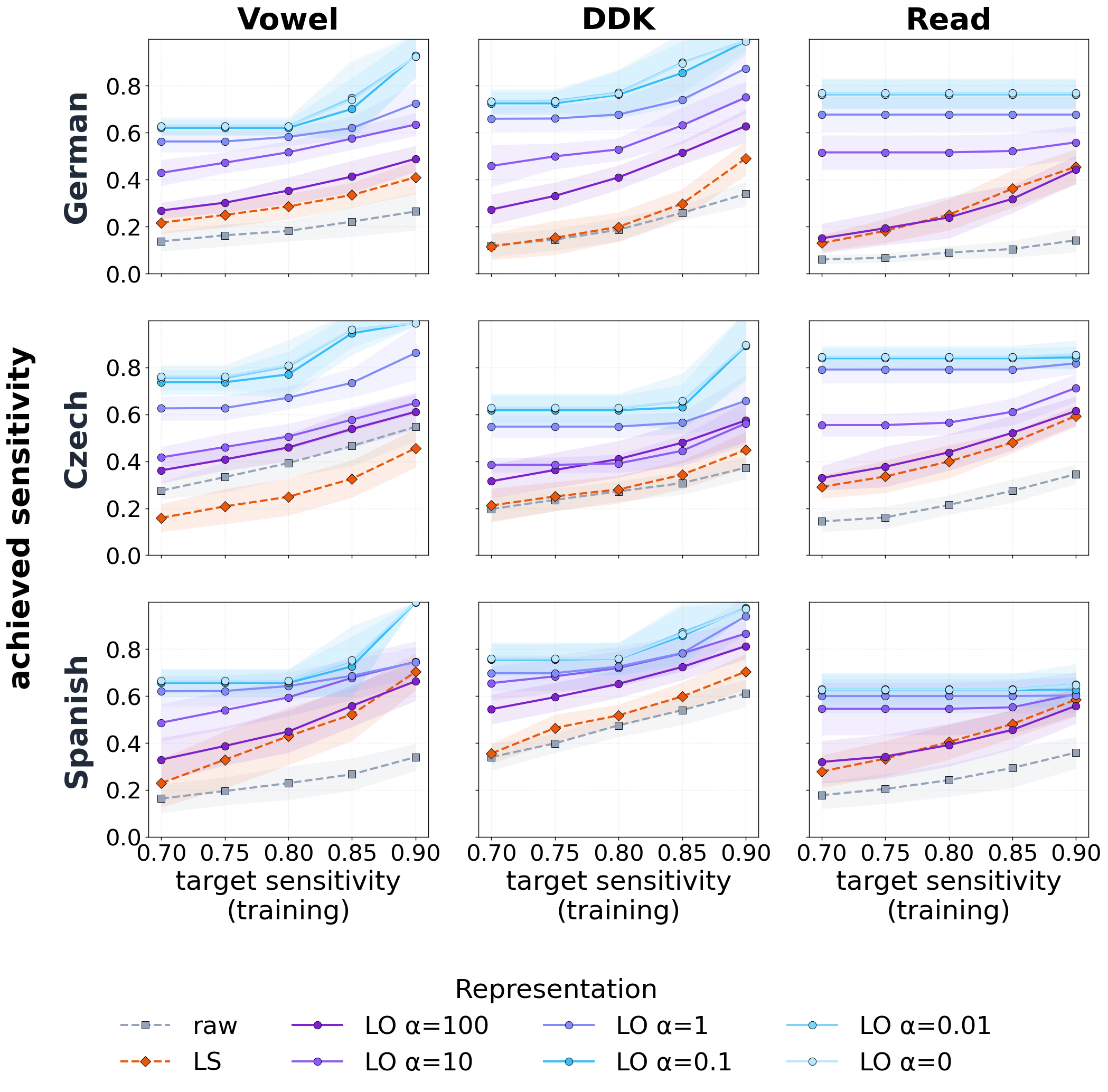}
  \caption{\textbf{Cross-lingual threshold transfer.}
  Each threshold is selected on the training set for a target sensitivity and then applied unchanged to the held-out target language. The diagonal indicates perfect agreement between the target and achieved sensitivities. Raw and LS fall well below the diagonal, whereas LO closely follows it across the $0.70$--$0.90$ range.}
  \label{fig:transfer}
\end{figure}

\begin{figure}[t]
  \centering
  \includegraphics[width=0.95\columnwidth]{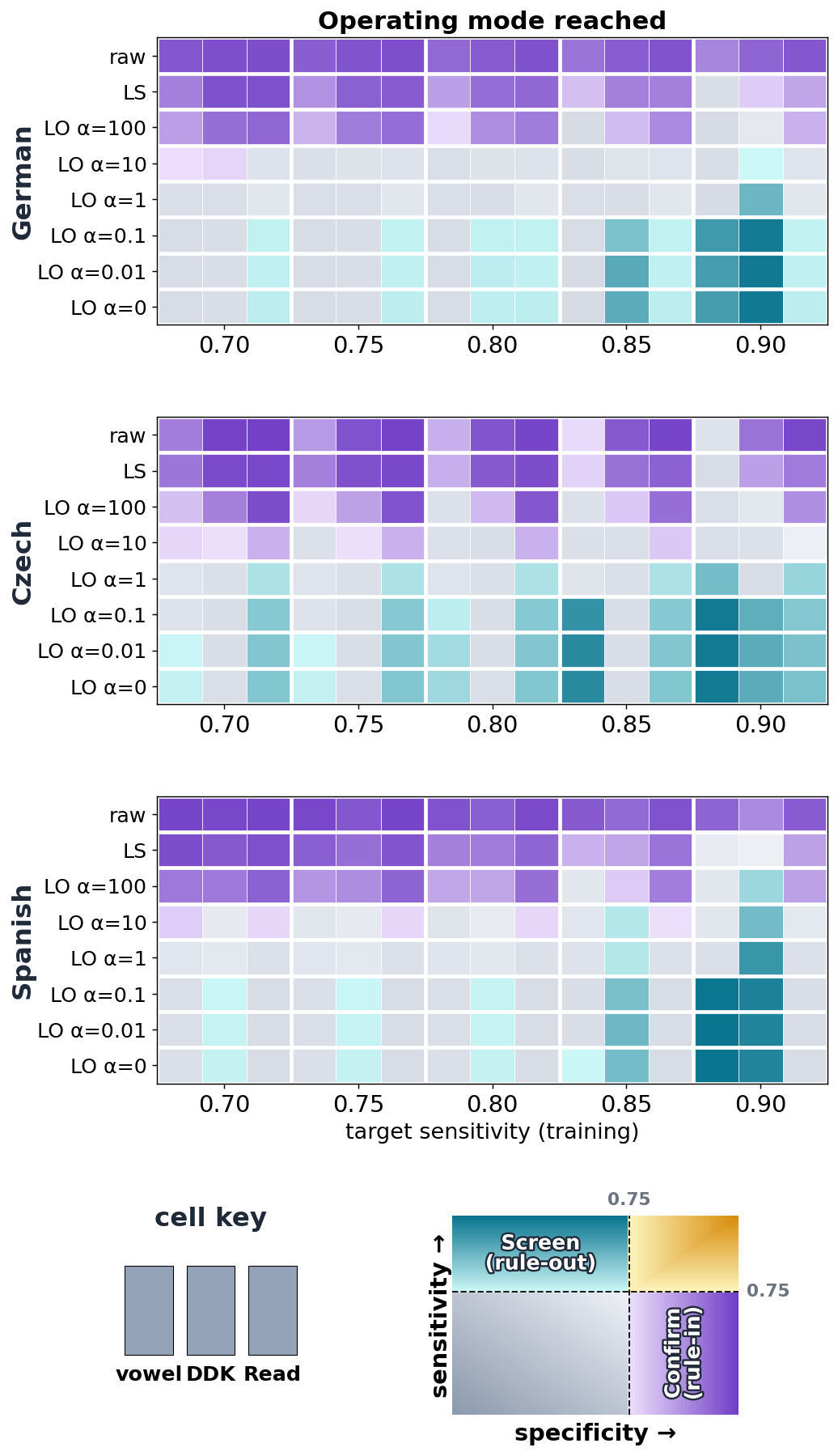}
  \caption{\textbf{Screening performance across target languages.}
  \emph{Screen} denotes sensitivity of at least $0.75$, whereas \emph{Confirm} denotes high specificity but sensitivity below $0.75$. Raw and LS remain in the \emph{Confirm} regime across target languages, while LO reaches the \emph{Screen} regime without target-language PD data during training.}
  \label{fig:opmode}
\end{figure}

\subsection{Cross-lingual PD detection}\label{ssec:sota}
\Cref{tab:f1_09} reports cross-lingual F1 at the $0.90$-sensitivity operating point, averaged over five S3M backbones and three target languages. With $\alpha=0$ used across all tasks and backbones, LO outperforms both baselines on every task, increasing average F1 from $0.52$ for raw features and $0.67$ for LS to $0.87$, a gain of $+0.35$ over raw features. \Cref{fig:f1_alpha} presents results across the full $\alpha$ sweep for all backbones and tasks. LO outperforms both baselines over a broad range of $\alpha$, with F1 generally increasing as $\alpha$ decreases, suggesting that removing more of the language-predictable component improves cross-lingual PD detection.

\subsection{Threshold transferability}\label{ssec:thr}
We use five-fold out-of-fold predictions on the training set to select a threshold for a requested sensitivity, then apply this threshold to the held-out target language. \Cref{fig:transfer} compares the requested sensitivity with the sensitivity achieved on target-language PD speakers. For raw and LS features, achieved sensitivity is consistently lower than requested, indicating that thresholds estimated from the training distribution do not transfer reliably. With LO, achieved sensitivity closely follows the requested sensitivity across the $0.70$--$0.90$ range. Thus, LO enables threshold selection without target-language PD data for calibration.

\subsection{Application to PD screening}\label{ssec:screen}
We next evaluate threshold transfer in a PD-screening setting. A screening (rule-out) test prioritizes sensitivity to minimize missed patients, whereas a confirmatory (rule-in) test prioritizes specificity~\cite{haynes1997evidence}. These objectives trade off through the decision threshold: lowering the threshold increases sensitivity at the cost of specificity, while raising it has the opposite effect.

Using the training-set thresholds selected in \Cref{ssec:thr}, we measure both sensitivity and specificity on each target language. As shown in \Cref{fig:opmode}, raw and LS remain in the high-specificity, low-sensitivity region despite thresholds selected for screening. LO yields substantially higher sensitivity and places the operating points in the screening region across all target languages. These results show that LO supports sensitivity-targeted screening without target-language PD data during training or threshold selection.

\begin{figure}[t]
  \centering
  \includegraphics[width=\columnwidth]{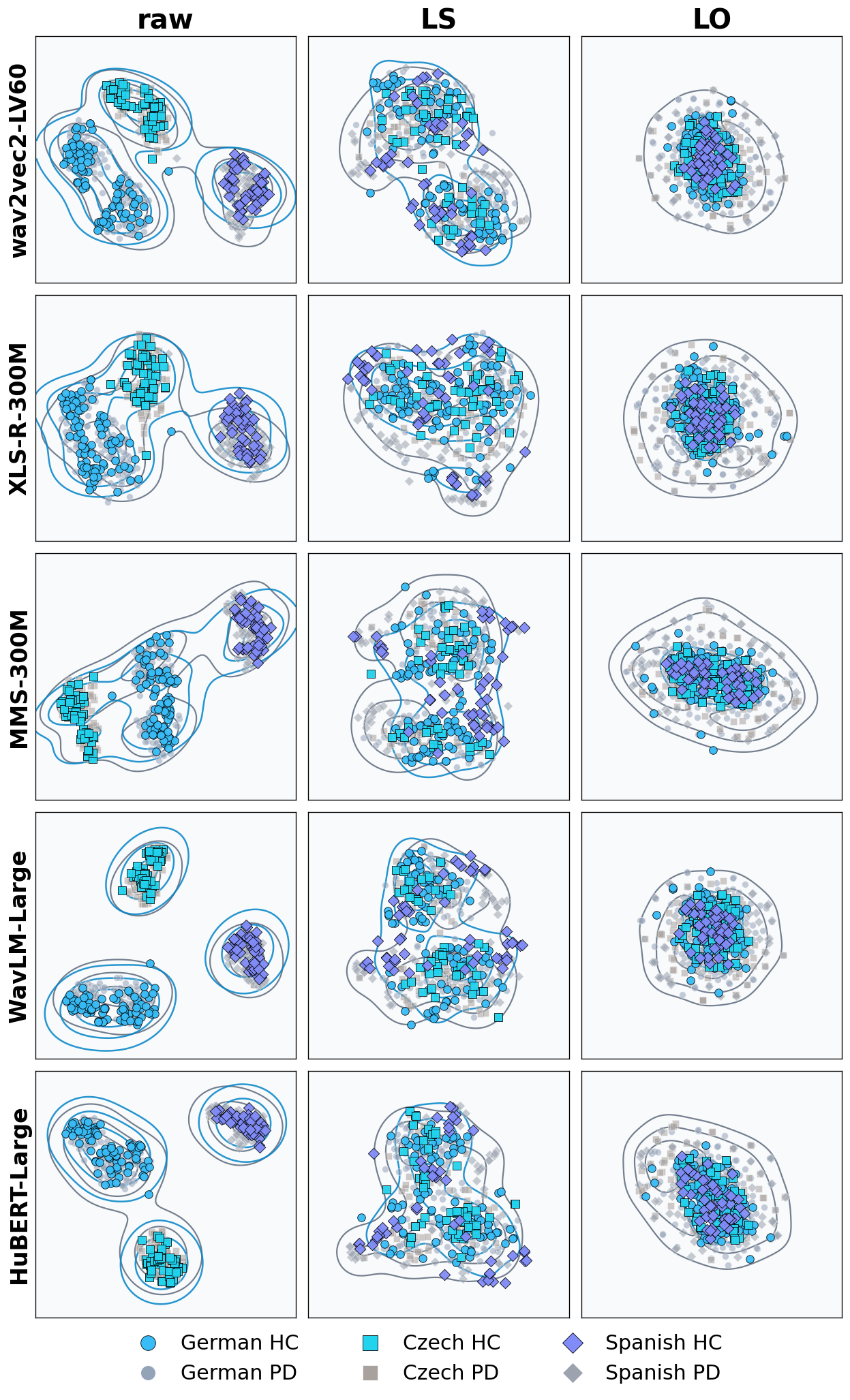}
  \caption{\textbf{t-SNE visualization before and after alignment.}
  An example from the read-speech task with German as the target language and $\alpha=0.01$. Raw representations separate strongly by language. LS brings the language clusters closer but leaves visible language-dependent structure, whereas LO produces greater overlap among the three languages. In the LO projection, HC representations are concentrated near the center and PD representations are more dispersed.}
  \label{fig:tsne}
\end{figure}
 
\begin{figure}[t]
  \centering
  \includegraphics[width=\columnwidth]{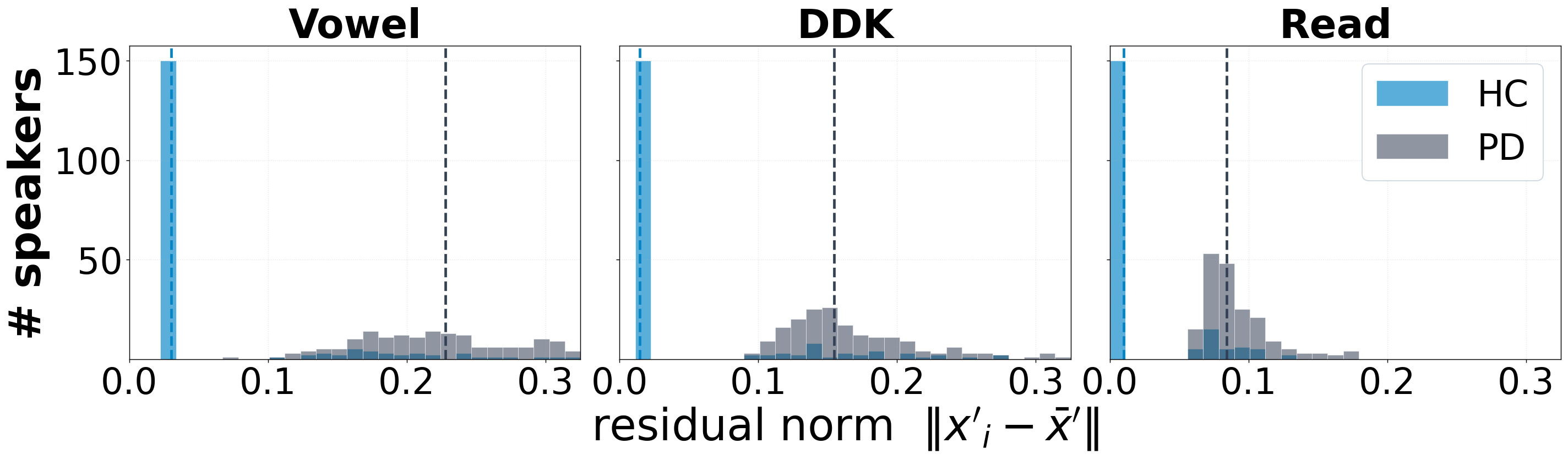}
  \caption{\textbf{Per-speaker residual norms after orthogonalization.} Norm $\lVert x'_i - \bar{x}'\rVert$ of each speaker's orthogonalized representation $x'_i$ about the mean residual $\bar{x}'$ (HuBERT-Large, $\alpha=0.01$). HC form a tight, low-norm baseline; PD spread to larger norms, with median PD norm $7.5$--$10.8\times$ the HC median across tasks, consistent with the heterogeneous nature of dysarthric deviation (\Cref{sec:mechanism}).}
  \label{fig:residual}
\end{figure}

\section{Analysis}
\label{sec:mechanism}

\subsection{Representation geometry}

\Cref{fig:tsne} illustrates how each transformation reorganizes the representation space, while \Cref{fig:residual} quantifies the resulting geometry using each speaker's distance from the residual centroid.

\noindent\textit{\textbf{Language shift: mean alignment.}}
LS (\Cref{eq:ls}) translates all representations from a language by the same vector. It therefore preserves the within-language covariance:
\begin{equation}
\Sigma'_L
= \mathrm{Cov}(\{x'_i:\ell_i{=}L\})
= \mathrm{Cov}(\{x_i:\ell_i{=}L\})
= \Sigma_L,
\label{eq:cov-inv}
\end{equation}
since $\mathrm{Cov}(x+c)=\mathrm{Cov}(x)$ for any constant $c$. Thus, LS aligns language centroids but does not remove language-dependent variation around them. The t-SNE projections in \Cref{fig:tsne} illustrate this distinction: LS brings the language clusters closer, while HC and PD remain largely intermixed.

\noindent\textit{\textbf{Language orthogonalization: residualization.}}
LO subtracts the speaker-specific prediction $W^*\tilde g_i$, rather than a single vector shared within each language. It can therefore remove variation associated with the LID embedding beyond differences in language means. In the t-SNE projections of \Cref{fig:tsne}, LO yields greater overlap among the three languages than LS. This qualitative pattern appears across the five backbones and three tasks.

\noindent\textit{\textbf{Residual geometry.}}
We quantify the resulting geometry using the Euclidean distance of each residual representation from the overall residual centroid, $\lVert x'_i-\overline{x}'\rVert_2$. HC speakers are concentrated near the centroid, whereas PD speakers are distributed farther from it. Across tasks, the median distance for PD speakers is $7.5$--$10.8$ times that for HC speakers (\Cref{fig:residual}). This pattern is consistent with the heterogeneous nature of dysarthric deviations in PD speech~\cite{duffy2012motor}.

\begin{table}[t]
\centering
\small
\setlength{\tabcolsep}{10pt}
\renewcommand{\arraystretch}{1.2}
\caption{\textbf{Language decodability after transformation.}
Macro-F1 for three-way language classification, averaged over five S3M
backbones and three speech tasks. Chance performance is $0.33$.}
\label{tab:lid_residual}
\begin{tabular}{l cc}
\toprule
Representation & HC & PD \\
\midrule
Raw & 0.99 & 0.98 \\
LS  & 0.76 & 0.94 \\
LO  & \textbf{0.47} & \textbf{0.62} \\
\bottomrule
\end{tabular}
\end{table}

\subsection{Residual language information}\label{ssec:lid}
To quantify the language information retained after each transformation, we train a linear classifier to predict the three languages from the resulting representations. We report speaker-disjoint five-fold macro-F1, averaged over five S3M backbones and three speech tasks, separately for HC and PD speakers. Chance performance is $0.33$.

As shown in \Cref{tab:lid_residual}, language is almost perfectly decodable from the raw representations, with macro-F1 scores of $0.99$ for HC and $0.98$ for PD. LS reduces language decodability for HC speakers to $0.76$, but little for PD speakers ($0.94$). LO yields the lowest language decodability for both groups, reducing macro-F1 to $0.47$ for HC and $0.62$ for PD. Thus, LO removes substantially more language information than LS, although some remains, particularly among PD speakers.

\subsection{Applicability beyond S3Ms}

\begin{figure}[t]
  \centering
  \includegraphics[width=\columnwidth]{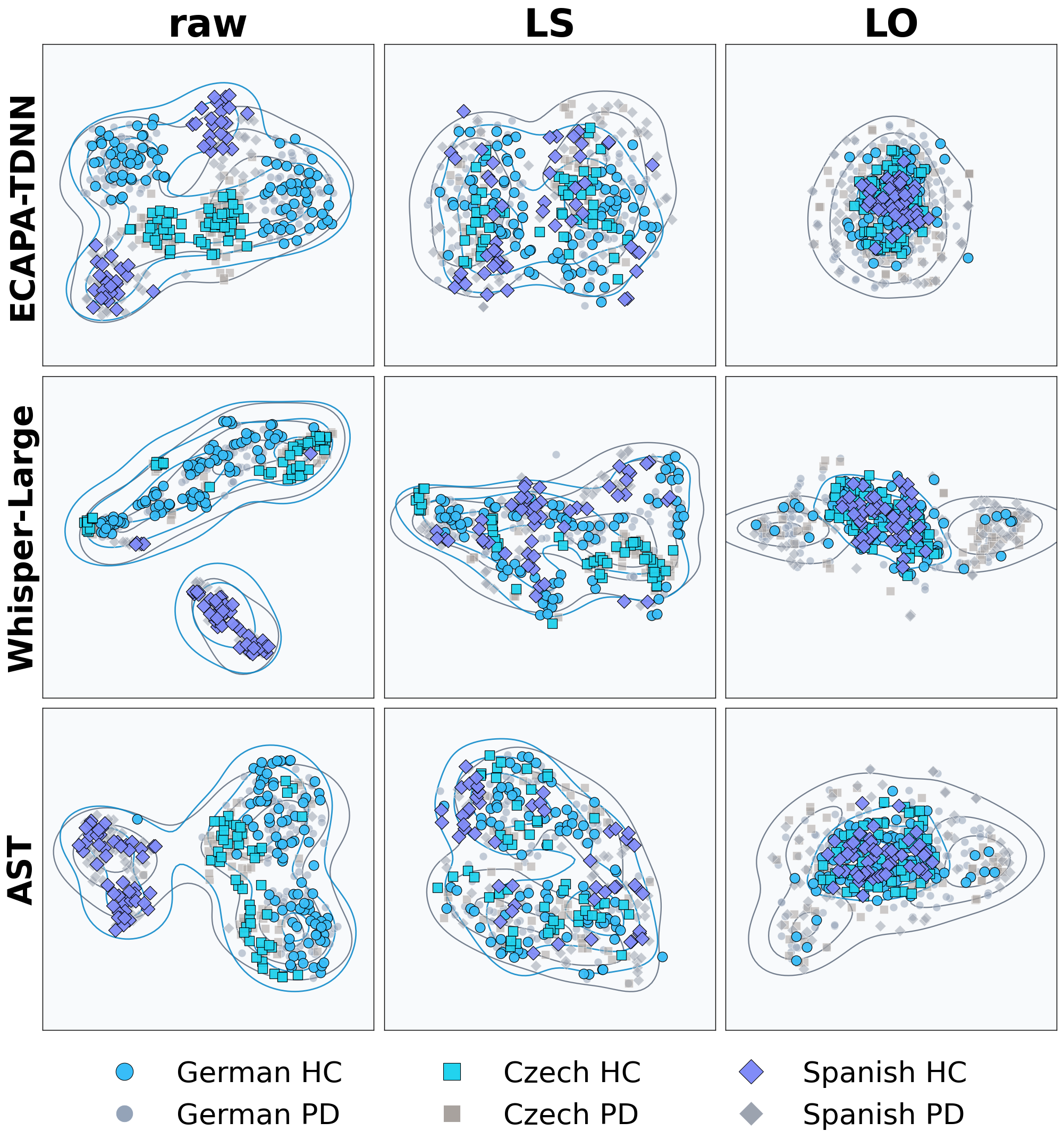}
  \caption{\textbf{Exploratory results on non-S3M representations.}
  t-SNE projections for the read-speech task with German as the target language and $\alpha=0.01$. Results are shown for a speaker embedding (ECAPA-TDNN), a supervised speech encoder (Whisper), and an audio transformer (AST) under raw, LS, and LO. Across these representations, LO produces greater overlap among languages while retaining variation between HC and PD speakers.}
  \vspace{20pt}
  \label{fig:generality}
\end{figure}

To explore the potential of LO beyond S3Ms, we apply it to three representations with different training objectives: a speaker embedding (ECAPA-TDNN~\cite{desplanques2020ecapa}), a supervised speech encoder (Whisper~\cite{radford2023robust}), and an audio transformer (AST~\cite{gong2021ast}). The t-SNE projections in \Cref{fig:generality} show a pattern similar to that observed for S3Ms: LO increases the overlap among languages while HC and PD speakers remain differently distributed. These exploratory findings motivate a more comprehensive evaluation of LO across other representation spaces.

\section{Discussion}
\label{sec:discussion}

\noindent\textit{\textbf{Controlling language information.}}
Language information can support cross-lingual transfer when used as an auxiliary signal~\cite{kim2025improving,khurana2024cross}. LO also leverages this information, but uses an external LID representation to identify and remove language-predictable variation rather than as an additional input to the downstream classifier. This distinction is important when target-language training data contain only HC speech, as the classifier may otherwise associate the target language with HC. The lower language decodability after LO supports this interpretation (\Cref{ssec:lid}).

\noindent\textit{\textbf{Generalization and interpretation.}}
Our evaluation covers Germanic, Slavic, and Romance languages, all within the Indo-European family. Extending it to more typologically diverse languages would provide a broader test of LO. S3M representation spaces tend to place similar languages closer together and dissimilar languages farther apart~\cite{kim2026scaling}, so LO may behave differently as the representational distance between source and target languages increases. Evaluating such language pairs is particularly important as S3Ms are increasingly applied cross-lingually~\cite{kim2026far,de2024layer}. LO also yields larger gains for sustained vowels than for read speech, indicating possible task dependence. Explainability analyses could clarify how language, task, and pathology cues interact in the removed and residual components~\cite{choi2025leveraging,yeo2023speech}. Similar analyses across other speech disorders would help establish the broader scope of LO.

\noindent\textit{\textbf{Methodological extensions.}}
LO is related to domain-adversarial learning with a gradient reversal layer (GRL)~\cite{ganin2016domain}. A future GRL baseline could treat language as the domain and reduce its predictability through adversarial fine-tuning, whereas LO operates on fixed representations through a single closed-form step. Recent non-linear modeling of pathological speech representations~\cite{kim2026hierarchical} also motivates extending LO beyond its current linear formulation. Building on our threshold-transfer analysis, future work could examine score calibration across languages to further support clinical deployment~\cite{brummer2006calibration}. Other directions include selecting $\alpha$ without target-language labels and improving specificity while maintaining screening sensitivity.

\section{Conclusion}
\label{sec:conclusion}
We presented language orthogonalization (LO) for cross-lingual Parkinson's disease detection, a closed-form, HC-only ridge residualization that uses external language embeddings to reduce language-predictable variation in S3M representations. Across five backbones, three speech tasks, and three target languages, LO consistently improved F1 over raw features and LS and mitigated the high-specificity, low-sensitivity failure on target-language PD speakers. It also improved the transfer of sensitivity-targeted thresholds without requiring target-language PD data, supporting cross-lingual screening. Language-decoding and representation analyses further showed that LO substantially reduces, although does not eliminate, language information. Exploratory results on non-S3M representations suggest that the approach may also extend to other speech representation spaces.


\section*{Disclosure of AI Usage}
Generative AI tools were used for minor language editing and coding support. All research design, analysis, and conclusions are the authors' own, and the authors take full responsibility for the content.
 
\bibliographystyle{IEEEtran}
\bibliography{mybib}

@article{hernandez2026adapting,
  title={Adapting Self-Supervised Speech Representations for Cross-lingual Dysarthria Detection in Parkinson's Disease},
  author={Hernandez, Abner and Yeo, Eunjung and Choi, Kwanghee and Li, Chin-Jou and Yue, Zhengjun and Das, Rohan Kumar and Rusz, Jan and Doss, Mathew Magimai and Orozco-Arroyave, Juan Rafael and Arias-Vergara, Tom{\'a}s and others},
  journal={arXiv preprint arXiv:2603.22225},
  year={2026}
}

@article{critchley1981speech,
  title={Speech disorders of Parkinsonism: a review.},
  author={Critchley, EM},
  journal={Journal of Neurology, Neurosurgery \& Psychiatry},
  volume={44},
  number={9},
  pages={751--758},
  year={1981},
  publisher={BMJ Publishing Group Ltd}
}

@article{yeo2025applications,
  title={Applications of artificial intelligence for cross-language intelligibility assessment of dysarthric speech},
  author={Yeo, Eunjung and Liss, Julie M and Berisha, Visar and Mortensen, David R},
  journal={Perspectives of the ASHA Special Interest Groups},
  volume={10},
  number={6},
  pages={2298--2308},
  year={2025},
  publisher={American Speech-Language-Hearing Association}
}

@inproceedings{li2025towards,
  title={Towards Inclusive ASR: Investigating Voice Conversion for Dysarthric Speech Recognition in Low-Resource Languages},
  author={Li, Chin-Jou and Yeo, Eunjung and Choi, Kwanghee and P{\'e}rez-Toro, Paula Andrea and Someki, Masao and Das, Rohan Kumar and Yue, Zhengjun and Orozco-Arroyave, Juan Rafael and N{\"o}th, Elmar and Mortensen, David R},
  booktitle={Proc. Interspeech 2025},
  pages={2128--2132},
  year={2025}
}

@inproceedings{yeo2022multilingual,
  title={Multilingual analysis of intelligibility classification using english, korean, and tamil dysarthric speech datasets},
  author={Yeo, Eun Jung and Kim, Sunhee and Chung, Minhwa},
  booktitle={2022 25th Conference of the Oriental COCOSDA International Committee for the Co-ordination and Standardisation of Speech Databases and Assessment Techniques (O-COCOSDA)},
  pages={1--6},
  year={2022},
  organization={IEEE}
}

@article{atalar2023hypokinetic,
  title={Hypokinetic Dysarthria in Parkinson's Disease: A Narrative Review.},
  author={Atalar, Merve Sapmaz and Oguz, Ozlem and Genc, Gencer},
  journal={Medical Bulletin of Sisli Etfal Hospital},
  volume={57},
  number={2},
  year={2023}
}

@article{darley1969differential,
  title={Differential diagnostic patterns of dysarthria},
  author={Darley, Frederic L and Aronson, Arnold E and Brown, Joe R},
  journal={Journal of speech and hearing research},
  volume={12},
  number={2},
  pages={246--269},
  year={1969},
  publisher={American Speech-Language-Hearing Association}
}

@book{duffy2012motor,
  title={Motor speech disorders: Substrates, differential diagnosis, and management},
  author={Duffy, Joseph R and others},
  year={2012},
  publisher={Elsevier Health Sciences}
}

@article{orozco2016automatic,
  title={Automatic detection of Parkinson's disease in running speech spoken in three different languages},
  author={Orozco-Arroyave, Juan Rafael and H{\"o}nig, F and Arias-Londo{\~n}o, JD and Vargas-Bonilla, JF and Daqrouq, K and Skodda, S and Rusz, J and N{\"o}th, E},
  journal={The Journal of the Acoustical Society of America},
  volume={139},
  number={1},
  pages={481--500},
  year={2016},
  publisher={AIP Publishing}
}

@article{lim2025cross,
  title={A cross-language speech model for detection of Parkinson’s disease: WS Lim et al.},
  author={Lim, Wee Shin and Chiu, Shu-I and Peng, Pei-Ling and Jang, Jyh-Shing Roger and Lee, Sol-Hee and Lin, Chin-Hsien and Kim, Han-Joon},
  journal={Journal of Neural Transmission},
  volume={132},
  number={4},
  pages={579--590},
  year={2025},
  publisher={Springer}
}

@article{thies2025automatic,
  title={Automatic speech analysis combined with machine learning reliably predicts the motor state in people with Parkinson’s disease},
  author={Thies, Tabea and Mallick, Elisa and Tr{\"o}ger, Johannes and Baykara, Ebru and M{\"u}cke, Doris and Barbe, Michael T},
  journal={npj Parkinson's Disease},
  volume={11},
  number={1},
  pages={105},
  year={2025},
  publisher={Nature Publishing Group UK London}
}

@article{cao2025speech,
  title={Speech and language biomarkers for Parkinson’s disease prediction, early diagnosis and progression},
  author={Cao, Fangyuan and Vogel, Adam P and Gharahkhani, Puya and Renteria, Miguel E},
  journal={npj Parkinson's Disease},
  volume={11},
  number={1},
  pages={57},
  year={2025},
  publisher={Nature Publishing Group UK London}
}

@article{rusz2024prodromal,
  title={From prodromal stages to clinical trials: The promise of digital speech biomarkers in Parkinson's disease},
  author={Rusz, Jan and Krack, Paul and Tripoliti, Elina},
  journal={Neuroscience \& Biobehavioral Reviews},
  volume={167},
  pages={105922},
  year={2024},
  publisher={Elsevier}
}

@article{kovac2025digital,
  title={Digital speech biomarkers for assessing cognitive decline across neurodegenerative conditions},
  author={Kovac, Daniel and Novakova, Lubomira and Mekyska, Jiri and Novotny, Krystof and Brabenec, Lubos and Klobusiakova, Patricia and Rektorova, Irena},
  journal={Computers in Biology and Medicine},
  volume={198},
  pages={111251},
  year={2025},
  publisher={Elsevier}
}

@article{hsu2021hubert,
  title={Hubert: Self-supervised speech representation learning by masked prediction of hidden units},
  author={Hsu, Wei-Ning and Bolte, Benjamin and Tsai, Yao-Hung Hubert and Lakhotia, Kushal and Salakhutdinov, Ruslan and Mohamed, Abdelrahman},
  journal={IEEE/ACM transactions on audio, speech, and language processing},
  volume={29},
  pages={3451--3460},
  year={2021},
  publisher={IEEE}
}

@article{chen2022wavlm,
  title={Wavlm: Large-scale self-supervised pre-training for full stack speech processing},
  author={Chen, Sanyuan and Wang, Chengyi and Chen, Zhengyang and Wu, Yu and Liu, Shujie and Chen, Zhuo and Li, Jinyu and Kanda, Naoyuki and Yoshioka, Takuya and Xiao, Xiong and others},
  journal={IEEE Journal of Selected Topics in Signal Processing},
  volume={16},
  number={6},
  pages={1505--1518},
  year={2022},
  publisher={IEEE}
}

@inproceedings{babu2022xls,
  title={XLS-R: Self-supervised Cross-lingual Speech Representation Learning at Scale},
  author={Babu, Arun and Wang, Changhan and Tjandra, Andros and Lakhotia, Kushal and Xu, Qiantong and Goyal, Naman and Singh, Kritika and von Platen, Patrick and Saraf, Yatharth and Pino, Juan and others},
  booktitle={Proc. Interspeech 2022},
  pages={2278--2282},
  year={2022}
}

@article{tsanas2012novel,
  title={Novel speech signal processing algorithms for high-accuracy classification of Parkinson's disease},
  author={Tsanas, Athanasios and Little, Max A and McSharry, Patrick E and Spielman, Jennifer and Ramig, Lorraine O},
  journal={IEEE transactions on biomedical engineering},
  volume={59},
  number={5},
  pages={1264--1271},
  year={2012},
  publisher={IEEE}
}

@inproceedings{bocklet2011detection,
  title={Detection of persons with Parkinson's disease by acoustic, vocal, and prosodic analysis},
  author={Bocklet, Tobias and N{\"o}th, Elmar and Stemmer, Georg and Ruzickova, Hana and Rusz, Jan},
  booktitle={2011 IEEE workshop on automatic speech recognition \& understanding},
  pages={478--483},
  year={2011},
  organization={IEEE}
}

@inproceedings{perez2021emotional,
  title={Emotional state modeling for the assessment of depression in Parkinson’s disease},
  author={P{\'e}rez-Toro, Paula Andrea and Vasquez-Correa, Juan Camilo and Arias-Vergara, Tom{\'a}s and Klumpp, Philipp and Schuster, Maria and N{\"o}th, Elmar and Orozco-Arroyave, Juan Rafael},
  booktitle={International Conference on Text, Speech, and Dialogue},
  pages={457--468},
  year={2021},
  organization={Springer}
}

@article{kim2017cross,
  title={A cross-language study of acoustic predictors of speech intelligibility in individuals with Parkinson's disease},
  author={Kim, Yunjung and Choi, Yaelin},
  journal={Journal of Speech, Language, and Hearing Research},
  volume={60},
  number={9},
  pages={2506--2518},
  year={2017},
  publisher={American Speech-Language-Hearing Association}
}

@article{kim2025cross,
  title={A Cross-Language Study of Oral Diadochokinesis: Rates and Rhythm},
  author={Kim, Yunjung and Berry, Jeffrey and Lee, Seung Jin and Lin, Lifeng},
  journal={Folia Phoniatrica et Logopaedica},
  year={2025}
}

@article{choi2026b,
  title={[b]=[d]-[t]+[p]: Self-supervised Speech Models Discover Phonological Vector Arithmetic},
  author={Choi, Kwanghee and Yeo, Eunjung and Cho, Cheol Jun and Harwath, David and Mortensen, David R},
  journal={arXiv preprint arXiv:2602.18899},
  year={2026}
}

@inproceedings{valk2021voxlingua107,
  title={VoxLingua107: a dataset for spoken language recognition},
  author={Valk, J{\"o}rgen and Alum{\"a}e, Tanel},
  booktitle={2021 IEEE Spoken Language Technology Workshop (SLT)},
  pages={652--658},
  year={2021},
  organization={IEEE}
}

@article{rios2024automatic,
  title={Automatic speech-based assessment to discriminate Parkinson’s disease from essential tremor with a cross-language approach},
  author={Rios-Urrego, Cristian David and Rusz, Jan and Orozco-Arroyave, Juan Rafael},
  journal={npj Digital Medicine},
  volume={7},
  number={1},
  pages={37},
  year={2024},
  publisher={Nature Publishing Group UK London}
}

@article{baevski2020wav2vec,
  title={wav2vec 2.0: A framework for self-supervised learning of speech representations},
  author={Baevski, Alexei and Zhou, Yuhao and Mohamed, Abdelrahman and Auli, Michael},
  journal={Advances in neural information processing systems},
  volume={33},
  pages={12449--12460},
  year={2020}
}

@article{pratap2024scaling,
  title={Scaling speech technology to 1,000+ languages},
  author={Pratap, Vineel and Tjandra, Andros and Shi, Bowen and Tomasello, Paden and Babu, Arun and Kundu, Sayani and Elkahky, Ali and Ni, Zhaoheng and Vyas, Apoorv and Fazel-Zarandi, Maryam and others},
  journal={Journal of Machine Learning Research},
  volume={25},
  number={97},
  pages={1--52},
  year={2024}
}

@article{klempir2026statistical,
  title={Statistical, multi-scale and attention-based layer pooling of Wav2Vec-2 speech embeddings for Parkinson's disease detection},
  author={Klempir, Ondrej and Mullerova, Juliana Grand and Krupicka, Radim},
  journal={Computers in Biology and Medicine},
  volume={200},
  pages={111368},
  year={2026},
  publisher={Elsevier}
}

@inproceedings{pasad2021layer,
  title={Layer-wise analysis of a self-supervised speech representation model},
  author={Pasad, Ankita and Chou, Ju-Chieh and Livescu, Karen},
  booktitle={2021 IEEE Automatic Speech Recognition and Understanding Workshop (ASRU)},
  pages={914--921},
  year={2021},
  organization={IEEE}
}

@inproceedings{pasad2023comparative,
  title={Comparative layer-wise analysis of self-supervised speech models},
  author={Pasad, Ankita and Shi, Bowen and Livescu, Karen},
  booktitle={ICASSP 2023-2023 IEEE International Conference on Acoustics, Speech and Signal Processing (ICASSP)},
  pages={1--5},
  year={2023},
  organization={IEEE}
}

@article{la2024exploiting,
  title={Exploiting foundation models and speech enhancement for Parkinson's disease detection from speech in real-world operative conditions},
  author={La Quatra, Moreno and Turco, Maria Francesca and Svendsen, Torbj{\o}rn and Salvi, Giampiero and Orozco-Arroyave, Juan Rafael and Siniscalchi, Sabato Marco},
  journal={arXiv preprint arXiv:2406.16128},
  year={2024}
}

@inproceedings{desplanques2020ecapa,
  title={ECAPA-TDNN: Emphasized Channel Attention, Propagation and Aggregation in TDNN Based Speaker Verification},
  author={Desplanques, Brecht and Thienpondt, Jenthe and Demuynck, Kris},
  booktitle={Proc. Interspeech 2020},
  pages={3830--3834},
  year={2020}
}

@inproceedings{radford2023robust,
  title={Robust speech recognition via large-scale weak supervision},
  author={Radford, Alec and Kim, Jong Wook and Xu, Tao and Brockman, Greg and McLeavey, Christine and Sutskever, Ilya},
  booktitle={International conference on machine learning},
  pages={28492--28518},
  year={2023},
  organization={PMLR}
}

@article{gong2021ast,
  title={AST: Audio Spectrogram Transformer},
  author={Gong, Yuan and Chung, Yu-An and Glass, James},
  journal={Interspeech 2021},
  year={2021},
  publisher={ISCA}
}

@article{haynes1997evidence,
  title={Evidence-based medicine: How to practice \& teach EBM},
  author={Haynes, R Brian and Sackett, David L and Richardson, W Scott and Rosenberg, William and Langley, G Ross},
  journal={Canadian Medical Association. Journal},
  volume={157},
  number={6},
  pages={788},
  year={1997},
  publisher={CMA Impact, Inc.}
}

@inproceedings{kim2025improving,
  title={Improving cross-lingual phonetic representation of low-resource languages through language similarity analysis},
  author={Kim, Minu and Jang, Kangwook and Kim, Hoirin},
  booktitle={ICASSP 2025-2025 IEEE International Conference on Acoustics, Speech and Signal Processing (ICASSP)},
  pages={1--5},
  year={2025},
  organization={IEEE}
}

@inproceedings{khurana2024cross,
  title={Cross-lingual transfer learning for low-resource speech translation},
  author={Khurana, Sameer and Dawalatabad, Nauman and Laurent, Antoine and Vicente, Luis and Gimeno, Pablo and Mingote, Victoria and Glass, James},
  booktitle={2024 IEEE International Conference on Acoustics, Speech, and Signal Processing Workshops (ICASSPW)},
  pages={670--674},
  year={2024},
  organization={IEEE}
}

@inproceedings{de2024layer,
  title={A layer-wise analysis of Mandarin and English suprasegmentals in SSL speech models},
  author={de la Fuente, Anton and Jurafsky, Dan},
  booktitle={Proc. Interspeech 2024},
  pages={1290--1294},
  year={2024}
}

@inproceedings{kim2026far,
  title={How Far Do SSL Speech Models Listen for Tone? Temporal Focus of Tone Representation under Low-Resource Transfer},
  author={Kim, Minu and Um, Ji Sub and Kim, Hoirin},
  booktitle={ICASSP 2026-2026 IEEE International Conference on Acoustics, Speech and Signal Processing (ICASSP)},
  pages={18297--18301},
  year={2026},
  organization={IEEE}
}

@article{choi2026self,
  title={Self-supervised speech models encode phonetic context via position-dependent orthogonal subspaces},
  author={Choi, Kwanghee and Yeo, Eunjung and Cho, Cheol Jun and Mortensen, David R and Harwath, David},
  journal={arXiv preprint arXiv:2603.12642},
  year={2026}
}

@article{kim2015automatic,
  title={Automatic intelligibility classification of sentence-level pathological speech},
  author={Kim, Jangwon and Kumar, Naveen and Tsiartas, Andreas and Li, Ming and Narayanan, Shrikanth S},
  journal={Computer speech \& language},
  volume={29},
  number={1},
  pages={132--144},
  year={2015},
  publisher={Elsevier}
}

@inproceedings{yeo2023automatic,
  title={Automatic severity classification of dysarthric speech by using self-supervised model with multi-task learning},
  author={Yeo, Eun Jung and Choi, Kwanghee and Kim, Sunhee and Chung, Minhwa},
  booktitle={ICASSP 2023-2023 IEEE International Conference on Acoustics, Speech and Signal Processing (ICASSP)},
  pages={1--5},
  year={2023},
  organization={IEEE}
}

@article{javanmardi2024pre,
  title={Pre-trained models for detection and severity level classification of dysarthria from speech},
  author={Javanmardi, Farhad and Kadiri, Sudarsana Reddy and Alku, Paavo},
  journal={Speech Communication},
  volume={158},
  pages={103047},
  year={2024},
  publisher={Elsevier}
}

@inproceedings{dibazar2002feature,
  title={Feature analysis for automatic detection of pathological speech},
  author={Dibazar, Alireza A and Narayanan, Shrikanth and Berger, Theodore W},
  booktitle={Proceedings of the second joint 24th annual conference and the annual fall meeting of the biomedical engineering society][engineering in medicine and biology},
  volume={1},
  pages={182--183},
  year={2002},
  organization={IEEE}
}

@inproceedings{dibazar2006pathological,
  title={Pathological voice assessment},
  author={Dibazar, Alireza A and Berger, Theodore W and Narayanan, Shrikanth S},
  booktitle={2006 international conference of the IEEE engineering in medicine and biology society},
  pages={1669--1673},
  year={2006},
  organization={IEEE}
}

@inproceedings{gupta2016pathological,
  title={Pathological speech processing: State-of-the-art, current challenges, and future directions},
  author={Gupta, Rahul and Chaspari, Theodora and Kim, Jangwon and Kumar, Naveen and Bone, Daniel and Narayanan, Shrikanth},
  booktitle={2016 IEEE international conference on acoustics, speech and signal processing (ICASSP)},
  pages={6470--6474},
  year={2016},
  organization={IEEE}
}

@article{sapkota2025all,
  title={Do all features matter? Layer-wise feature probing of self-supervised speech models for dysarthria severity classification},
  author={Sapkota, Paban and Srivastava, Harsh and Kathania, Hemant Kumar and Narayanan, Shrikanth and Kadiri, Sudarsana Reddy},
  journal={Speech Communication},
  pages={103326},
  year={2025},
  publisher={Elsevier}
}

@article{maxim2014screening,
  title={Screening tests: a review with examples},
  author={Maxim, L Daniel and Niebo, Ron and Utell, Mark J},
  journal={Inhalation toxicology},
  volume={26},
  number={13},
  pages={811--828},
  year={2014},
  publisher={Taylor \& Francis}
}

@article{bone2016use,
  title={Use of machine learning to improve autism screening and diagnostic instruments: Effectiveness, efficiency, and multi-instrument fusion},
  author={Bone, Daniel and Bishop, Somer L and Black, Matthew P and Goodwin, Matthew S and Lord, Catherine and Narayanan, Shrikanth S},
  journal={Journal of Child Psychology and Psychiatry},
  volume={57},
  number={8},
  pages={927--937},
  year={2016},
  publisher={Wiley Online Library}
}

@inproceedings{choi2025leveraging,
  title={Leveraging allophony in self-supervised speech models for atypical pronunciation assessment},
  author={Choi, Kwanghee and Yeo, Eunjung and Chang, Kalvin and Watanabe, Shinji and Mortensen, David R},
  booktitle={Proceedings of the 2025 Conference of the Nations of the Americas Chapter of the Association for Computational Linguistics: Human Language Technologies (Volume 1: Long Papers)},
  pages={2613--2628},
  year={2025}
}

@inproceedings{yeo2023speech,
  title={Speech Intelligibility Assessment of Dysarthric Speech by using Goodness of Pronunciation with Uncertainty Quantification},
  author={Yeo, Eun Jung and Choi, Kwanghee and Kim, Sunhee and Chung, Minhwa},
  booktitle={Proc. Interspeech 2023},
  pages={166--170},
  year={2023}
}

@article{kim2026hierarchical,
  title={A Hierarchical Feature Engineering Framework for Automated Classification of Phonotraumatic and Non-Phonotraumatic Vocal Hyperfunction},
  author={Kim, June-Woo and Jang, Kangwook and Kim, Minu and Lee, Hyunju},
  journal={arXiv preprint arXiv:2606.07673},
  year={2026}
}

@inproceedings{brummer2006calibration,
  title={On calibration of language recognition scores},
  author={Brummer, Niko and Van Leeuwen, David A},
  booktitle={2006 IEEE Odyssey-The Speaker and Language Recognition Workshop},
  pages={1--8},
  year={2006},
  organization={IEEE}
}

@article{ganin2016domain,
  title={Domain-adversarial training of neural networks},
  author={Ganin, Yaroslav and Ustinova, Evgeniya and Ajakan, Hana and Germain, Pascal and Larochelle, Hugo and Laviolette, Fran{\c{c}}ois and March, Mario and Lempitsky, Victor},
  journal={Journal of machine learning research},
  volume={17},
  number={59},
  pages={1--35},
  year={2016}
}

@article{kim2026scaling,
  title={Scaling Self-Supervised Speech Models Uncovers Deep Linguistic Relationships: Evidence from the Pacific Cluster},
  author={Kim, Minu and Kim, Hoirin and Mortensen, David R},
  journal={arXiv preprint arXiv:2603.07238},
  year={2026}
}
 
\end{document}